\documentstyle[preprint,revtex]{aps}

\begin{document}

\begin{flushright}
{\large LBL-34156}
\end{flushright}

\draft
\begin{title}
Space-time Structure of Initial Parton Production in\\
Ultrarelativistic Heavy Ion Collisions
\end{title}
\author{K. J. Eskola$^{1,2}$ and Xin-Nian Wang$^1$}
\begin{instit}
$^1$ Nuclear Science Division, Mailstop 70A-3307,
			Lawrence Berkeley Laboratory\\
University of California, Berkeley, California 94720. \\
$^2$ Laboratory of High Energy Physics, P.O. Box 9 \\
SF-00014 University of Helsinki, Finland.
\end{instit}
\receipt{ }
\begin{abstract}

	The space and time evolution of initial parton production
in ultrarelativistic heavy ion collisions is investigated within
the framework of perturbative QCD which includes both initial and
final state radiations. Uncertainty principle is used to relate
the life time of a radiating parton to its virtuality and momentum.
The interaction time of each hard or semihard parton scattering is
also taken into account. For central $Au+Au$ collisions at
$\sqrt{s}$=200 GeV, most of the partons are found to be produced
within 0.5 fm/c after the total overlap of the two
colliding nuclei. The local momentum distribution is approximately
isotropical at that time. The implication on how to treat correctly
the the secondary scattering in an ultimate parton cascading model
is also discussed.

\end{abstract}
\narrowtext

\baselineskip=18pt

\section{INTRODUCTION}

	In the search for a quark gluon plasma (QGP), a thermally
and chemically equilibrated system of quarks and gluons freed from
the color confinement, heavy nuclei are
accelerated to collide with each other at ultrarelativistic
energies. Studies in the last few years \cite{KAJA}-\cite{RANF}
suggest that heavy ion
collisions are dominated by minijet production via semihard
parton scatterings with $p_T\sim$ few GeV/c at energies of
the Relativistic Heavy Ion Collider (RHIC) of Brookhaven National
Laboratory (BNL) and the proposed Large Hadron Collider (LHC)
of CERN. These minijets may contribute to most of the transverse
energy produced in the heavy ion collisions. If there is enough
time and if the density is sufficiently high for the partons
to rescatter, the initially
produced minijets will eventually lead to a thermalized QGP.
However, in order to estimate the initial energy density of the
produced partons, one must know their formation time. The total
formation time must include the interaction time of the semihard
processes and the formation time of initial and final state radiation.

	Another motivation of our study in this paper is to
understand the space-time evolution of the initial parton
production and the consequences on secondary scatterings.
One has to consider these scatterings in order to simulate the
thermalization process by parton cascading. Although there are
already Monte Carlo simulations of parton cascading in heavy ion
collisions \cite{KGBM,RANF}, the possibility of double counting
is still a potential problem.
One type of double counting may occur if one neglects
the interaction time of the scatterings. During the interaction
time, which depends on whether the scatterings is hard, semihard
or soft, the participating and therefore the produced partons
cannot scatter again immediately with the beam partons.
Another double counting has to do with the initially radiated
partons before the hard or semihard scatterings. In calculating
jet cross sections and the number of hard scatterings, parton
structure functions of a nucleus $f_{a/A}(x,Q^2)$ evaluated
at the scale $Q^2=p_T^2$ are used. These parton distribution
functions then have already included the effect of QCD evolution,
producing more partons at small $x$ and larger $Q^2$. Therefore,
in Monte Carlo simulations in which initial radiations are treated
as backward evolution, the radiated partons should not participate
in any interactions before the end of the corresponding hard scattering.
This is also in coincidence with the factorization theorem \cite{COLL}
of perturbative QCD (pQCD).

	In this paper, we will use HIJING Monte Carlo model \cite{WANG},
which was developed for simulating parton and particle production in heavy
ion collisions, to study the space-time evolution of the initial
parton production. In HIJING, each hard parton scattering in hadronic
reactions is simulated by using some subroutines of PYTHIA \cite{PYTH}
in which one can trace back the whole history of initial and final state
radiation associated with that hard scattering. The life time of
an intermediate off-shell parton is estimated via the
uncertainty principle $t_q\sim E/q^2$ for given energy $E$ and
virtuality $q$. We then obtain the space and time vertices of
all the produced partons. Given a space cell, we  then
calculate the time evolution of the momentum distribution of the
produced partons by assuming classical trajectories without rescatterings.
We will verify that after some time most of the
partons will leave the space cell except those whose rapidities
are close to the spatial rapidity of the cell, thus achieving local
isotropy in momentum distribution which happens to look like thermal.
This will provide a good starting point for treating equilibration
via a set of rate equations as in Refs. \cite{BDMTW,KEMG} for the
number and energy densities. However, what is not studied in this
paper, though included in HIJING, is the particle production from
the soft component.

	This paper is organized as follows: In Sec. I, we briefly
review the HIJING model and the related aspects of parton shower
in PYTHIA. More detailed descriptions can be found in Refs. \cite{WANG,PYTH}.
In Sec. II, we discuss how we implement the Lorentz contracted spatial
parton distribution of a nucleus into the calculation and how
we estimate the life time of a virtual parton and the interaction
time of semihard processes. The results are presented
and discussed in Sec. III, where consequences for secondary
parton scattering and thermalization processes are also discussed.
We give our conclusions in Sec. IV.

\section{HIJING MODEL}

	To make a consistent study of the initial parton production,
we briefly review the HIJING model and the structure of initial and
final state radiation in PYTHIA which HIJING model utilizes exclusively
for hard parton scatterings and the associated radiations. Readers
can find detailed descriptions in Refs. \cite{WANG,PYTH}.

\subsection{Minijet Production}

	In perturbative QCD, given the parton structure
functions $f_{a/N}(x,Q^2)$ and the perturbative parton-parton
cross section $d\sigma_{ab}$, the differential cross section for
jet production in nucleon-nucleon collisions can be calculated
via \cite{EHLQ}
\begin{equation}
	\frac{d\sigma_{\rm jet}}{dp_T^2dy_1dy_2} = K
	\sum_{\stackrel{ab}{cd}}
	x_1f_{a/N}(x_1,p_T^2)x_2f_{b/N}(x_2,p_T^2)
	\frac{d\sigma}{d\hat{t}}^{ab\rightarrow cd},
	\label{eq:sjet}
\end{equation}
where $x_{1,2}$ are the fractional momenta of the colliding
partons, $y_{1,2}$ are the rapidities of the jets and $p_T$ is
their transverse momentum. The factor $K\approx 2$ accounts roughly
for the corrections beyond the leading order.
We want to emphasize here that the jet cross sections are calculated
with the parton structure functions evaluated at the scale of
the hard scattering $Q^2=p_T^2$. These structure
functions $f_{a/N}(x,Q^2)$ are obtained by evolving them
from a lower scale  $Q_0^2$ via Altarelli-Parisi equations \cite{ALPA}
thus increasing the parton  density at lower values of $x$.

	Given the integrated inclusive jet cross section $\sigma_{\rm jet}$,
we can use the unitarized eikonal model \cite{CAPE}-\cite{GAIS}
to calculate the total nucleon-nucleon cross section and the
cross sections for multiple independent jet production \cite{WANG91a}
\begin{eqnarray}
	\sigma_0 &=& \int d^2b \left[1-e^{-\sigma_{\rm soft}T_N(b)}\right]
			e^{-\sigma_{\rm jet}T_N(b)}, \label{eq:sjt0}\\
	\sigma_j &=& \int d^2b \frac{[\sigma_{\rm jet}T_N(b)]^j}{j!}
			e^{-\sigma_{\rm jet}T_N(b)},\label{eq:sjtj}
\end{eqnarray}
where $T_N(b)$ is the partonic overlap function between two colliding
nucleons at impact parameter $b$.

	In this model two phenomenological parameters have to be
introduced \cite{WANG91a}: An infrared cutoff $p_0=2$ GeV/c is
used to calculate
the total inclusive ``hard'' parton interaction cross section
$\sigma_{\rm jet}(p_0,\sqrt{s})$. Another parameter,
$\sigma_{\rm soft}=57$ mb, is used to
characterize the corresponding ``soft'' parton interactions.
Most of the differences among several existing models \cite{WANG,KGBM,RANF}
result
from different values of these two parameters used.
As demonstrated in Ref. \cite{WANG91a}, these two
parameters, though constrained
by the total $pp$ and $p\bar{p}$ cross sections, are still
model dependent. Since most of the minijets are nonresovable
as distinct hadronic clusters in the calorimeter of an
experimental detector, their existence can only be justified
by their contribution to particle production in low and
intermediate $p_T$ region. Therefore, the value of $p_0$ and
$\sigma_{\rm soft}$ will depend on what is included in the
particle production from the so-called ``soft'' interactions.
Recently, two-particle correlation functions in azimuthal
angle and their energy and $p_T$ dependence are
proposed \cite{WANG92} to give further constraints
on these two parameters.

	Extrapolated to heavy ion collisions which are decomposed
into binary nucleon-nucleon scatterings, Eqs.(\ref{eq:sjt0},\ref{eq:sjtj})
are used to determine the number of minijet production in each
binary collision. Each hard scattering is then simulated
via subroutines of PYTHIA and initial and final state radiations
are generated. At ultrarelativistic energies, the colliding
nuclei are highly contracted longitudinally in the center-of-mass
frame but are surrounded by soft partons with the same longitudinal size
as a nucleon. The hard or semihard scatterings all happen at
around the same time when the two nuclei pass through each other.
Therefore, binary collision is a good approximation for hard
scatterings. In each binary collision the same parton structure
can be used. In other words, each individual nucleon really
does not have time to readjust its parton distribution before
the next binary collision. Of course, in the actual simulation,
energy and momentum conservation is strictly imposed.
To take into account the nuclear modification of the parton
structure functions, we parameterize \cite{WANG} the nuclear
shadowing effect as measured experimentally in deeply inelastic
lepton-nucleus scatterings and assume gluons and quarks are
shadowed by the same amount in small $x$ region. The inclusion
of nuclear shadowing will effectively reduce the number
of minijet production by about \%50 at RHIC energy. However,
considering the QCD evolution of the shadowing \cite{ESKO} may
reduce the effective shadowing.

\subsection{Initial and Final State Radiation}

	For each hard scatterings, one then has to take
into account the corrections due to initial and final
state radiations.  In an axial gauge and to the leading pole
approximation, the interference terms of the radiation
drop out. The amplitude for successive radiations has then a
simple ladder structure and the probability for multiple
emission becomes the product of each emission \cite{FIED}.
The virtualities of the radiating partons are ordered along
the tree, decreasing until a final value $\mu_0^2$ below which pQCD
is no longer valid any more. This provides a framework for a
Monte Carlo simulation of parton shower and its space-time
interpretation \cite{ODOR,WEBB}.

	At a given vertex of the branching tree, the probability
for the off-shell parton $a$ of virtuality $q^2<q^2_{\rm max}$
to branch into partons $b$ and $c$ with fractions $z$ and $1-z$
of the light-cone momentum is given by \cite{ODOR,WEBB}

\begin{equation}
	d{\cal P}_{a\rightarrow bc}(q^2,z)= \frac{dq^2}{q^2}
	dz\,P_{a\rightarrow bc}(z)\frac{\alpha_s[z(1-z)q^2]}{2\pi}
	\frac{{\cal S}_a(q^2_{\rm max})}{{\cal S}_a(q^2)},
		 \label{eq:shr1}
\end{equation}
where $P_{a\rightarrow bc}(z)$ is the Altarelli-Parisi
splitting function \cite{ALPA} for $a\rightarrow bc$ process.
By requiring the relative transverse momentum $q_T$ of $b$ and $c$
to be real,
\begin{equation}
	q_T^2=z(1-z)\left(q^2-\frac{q_b^2}{z}-\frac{q_c^2}{1-z}
	\right)\geq 0, \label{eq:shr2}
\end{equation}
and a minimum virtuality $q_b^2,\,q_c^2\geq\mu_0^2$,
the kinematically allowed region of phase space is
then,
\begin{eqnarray}
	4\mu_0^2< & q^2 & <q^2_{\rm max}; \nonumber \\
	\epsilon(q)< & z & <1-\epsilon(q),\ \
	\epsilon(q)=\frac{1}{2}(1-\sqrt{1-4\mu_0^2/q^2}).
\end{eqnarray}
The Sudakov form factor ${\cal S}_a(q^2)$ is defined as
\begin{equation}
	{\cal S}_a(q^2)=\exp\left\{
	-\int_{4\mu_0^2}^{q^2}\frac{dk^2}{k^2}
	 \int_{\epsilon(k)}^{1-\epsilon(k)}dz\sum_{b,c}
	P_{a\rightarrow bc}(z)\frac{\alpha_s[z(1-z)k^2]}{2\pi}
	\right\}, \label{eq:sdk}
\end{equation}
so that ${\cal S}_a(q^2_{\rm max})/{\cal S}_a(q^2)$ is
the probability for parton $a$ not to have any branching
between $q^2_{\rm max}$ and $q^2$.

	In principle, one can perform the initial state
radiation in a similar way. The partons inside a nucleon
can initiate a space-like branching increasing their
virtuality from some initial value $Q_0^2$. A hard scattering
can be considered as a probe which can only resolve
partons with virtuality up to the scale of the hard scattering.
Otherwise without the scattering, the off-shell partons are only
virtual fluctuations inside the hadron and they will reassemble
back to the initial partons. In PYTHIA, which uses backward evolution,
a hard scattering is selected first with the known QCD-evolved
structure function at that scale, and then the initial branching
processes are reconstructed down to the initial scale $Q_0^2$. The
evolution equations are essentially the same as in final state
radiation except that one has to convolute with the parton
structure functions. Readers can find details in Ref. \cite{PYTH}.

\section{FORMATION AND INTERACTION TIME}

	When a parton is off-shell, it can be considered as
a virtual fluctuation and it can only live for a finite
time, $\Delta t$, determined by its virtuality $q^2$ via the
uncertainty principle,
\begin{equation}
	\Delta t\approx q_0/q^2, \label{eq:dt}
\end{equation}
where $q_0$ is the energy of the parton. After $\Delta t$, the
off-shell parton will then branch or ``decay'' into other
partons which can further initiate branchings until a minimum
virtuality $\mu_0$ is reached. At $q^2\leq\mu_0^2$, pQCD is not
considered to be valid anymore and the process of nonperturbative
hadronization
takes over. Following this tree of branching (which also includes
initial space-like radiation) and assuming a straight line trajectory
for partons, we can then calculate the space and time evolution
of the initial parton production. Since we don't consider
secondary parton-parton interactions, the produced final partons
which are on shell are considered free particles. Fig.{}~\ref{fig1}
is a demonstration
of a hard scattering and the associated  branching trees.
The number on each line represents the life time (in fm/c)
of that virtual parton and the numbers in
parentheses are the corresponding virtualities (in GeV).

	The life time of a virtual parton in Eq.{}~\ref{eq:dt} is
only a rough estimate according to the uncertainty principle. One
could also estimate it via pQCD in lowest order \cite{KGBM} which
gives an extra factor of $1/\alpha_s$. But higher order corrections
could easily cancel this factor, so that $q_0/q^2$ should be a
good estimate of $\Delta t$ in magnitude.

	We also use Eq.{}~\ref{eq:dt} to estimate the interaction
time for each hard scattering with $q$ being the sum of the initial
or final four-momentum of the colliding partons. If the fractional
momenta of the partons are $x_1$ and $x_2$, then the interaction
time can be estimated as
\begin{equation}
		\Delta t_i\approx \frac{x_1+x_2}{2\,x_1\,x_2\sqrt{s}}.
\end{equation}
In this case, the asymmetric scatterings ($x_1\gg x_2$ or $x_1\ll x_2$)
have longer interaction time than the symmetric ones ($x_1\sim x_2$)
for fixed parton-parton center-of-mass energy $x_1\,x_2\sqrt{s}$.
We also assume that the interaction time is the same for all
channels.

	In the rest frame of each nucleus, three-parameter Wood-Saxon
nuclear densities are used to construct the nucleon distribution
inside the nucleus. The system is then boosted to the
center-of-mass frame of the two colliding nuclei.
Due to the fact that gluons, sea quarks and antiquarks are
only quantum fluctuations before they really
suffer scatterings, their longitudinal distribution
around the center of the nucleon is still governed by
the uncertainty principle in any boosted frame \cite{AM88}.
We refer to this distribution as the ``contracted distribution'', in
which a parton with $x_i$ fractional momentum has a finite
spatial spread,
\begin{equation}
	\Delta z_i\approx 2/x_i\sqrt{s}. \label{eq:dz}
\end{equation}
Transversely, partons are distributed around their parent
nucleons according to the Fourier
transform of a dipole form factor \cite{WANG91a}.
If we define $t=0$ as the moment when the two nuclei have complete
overlap, then the interaction point of two partons in a
$t-z$ plane can be anywhere within the shaded area in Fig.{}~\ref{fig2}.
The solid lines are the trajectories of the two parent
nucleons which spread around the nuclei according to a
longitudinally contracted Wood-Saxon distribution.

	Because of the spread of parton interaction points
in space and time, it is ambiguous to define a common
proper time in the $t-z$ plane for parton production in
heavy ion collisions. Therefore, as we will demonstrate in the next
section, a proper time,
\begin{equation}
	\tau=\sqrt{t^2-z^2}, \label{eq:tau}
\end{equation}
defined with respect to the nucleus overlap point, is
only approximately relevant.

\section{NUMERICAL RESULTS}

	In this section, we present the numerical results of our
study. All our results are for central $Au+Au$ collisions at
RHIC energy. In our simulation, Duke-Owens parton
distributions \cite{DKOW} for nucleons are used. The initial
virtuality for the initial state evolution is
set to be $Q_0=$2 GeV/c, and the
minimum virtuality for the final state radiation is $\mu_0=$ 0.5 GeV/c.
The maximum virtuality for the associated radiations in a
hard scattering with transverse momentum transfer $p_T$
is chosen to be $q_{\rm max}=2p_T$.
In PYTHIA, angular ordering is also enforced to take into
account the soft gluon interference \cite{WEBB} in the final state
radiation. In HIJING model, soft interactions as string excitations
are also included. These soft
interactions must also consume energy and affect minijet
production slightly. We want to emphasize that in our calculation here,
we do {\it not} include the soft interactions, so the number of
minijet production
here should be slightly more than the full calculation when
soft interactions are included.

\subsection{Space-time Evolution}

	To estimate the parton production time, we plot in
Fig.{}~\ref{fig3} the total number of produced partons, on-shell
as well as off-shell, as a function of time at the highest
RHIC energy, $\sqrt{s}=200$ GeV/n.
We see that long  before the two nuclei overlap and
hard scatterings take place, partons have already
been produced via initial state bremsstrahlung.
Some of the initially radiated partons will also
initiate time-like branching trees(see Fig.\ref{fig1}).
Note that, if the coherence is not broken by the hard
scattering, partons which would have been emitted from
the initial state radiation will not emerge as produced
partons. Here we have also included the initiators of the
space-like branching as produced partons. Therefore, if
a parton does not have initial state radiation, it will only
become a produced parton after the hard scattering, whereas,
a parton is defined to be produced before the hard
scattering if it has initial state radiation. {}From Fig.{}~\ref{fig3},
we can see that about $2/3$ of the total number of partons
are produced between $t=-0.5$ and 0.5 fm/c while about $200$
semihard scatterings happen between $t=-0.1$ and 0.1 fm/c as
indicated by dashed lines. We find also that about $2/3$ of the
total number of partons are produced in initial and final
state radiations. The fraction of partons from branching
should increase with the colliding energy and with smaller
choices for $\mu_0$.

	To see how hard or semihard scatterings and initial
and final state bremsstrahlung contribute to parton
production, we show in Fig.{}~\ref{fig4} the rapidity distribution
of produced partons at different time. Before $t=0$,
most of the partons come from initial state radiation.
Since the radiations are almost collinear, these partons move
along the beam direction and therefore have large rapidities.
The semihard scatterings then produce partons uniformly over
a rapidity plateau and fill up the middle rapidity region.
Final state radiations, which happen after the semihard
scatterings, will also produce partons uniformly in the
central rapidity region.  We therefore see from Fig.{}~\ref{fig4}
that the dip of $dN/dy$ in middle rapidity is actually
caused by the parton production from initial state
bremsstrahlung at large rapidity. These partons
with large longitudinal momenta will move away
from the interaction region after the semihard scatterings.
They do not rescatter with the beam partons in the
leading twist approximation. We will come back to
this point again when we discuss the consequences of our
study on how to treat parton rescattering. In Fig.{}~\ref{fig5},
we also show the time evolution of the $p_T$ distribution of
the produced partons. Since the partons, which initially
have a Gaussian $p_T$ distribution, have gone through
initial state radiation, they have already a large $p_T$
tail at $t=-0.6$ fm/c before the hard scattering. Hard
scatterings will transfer large transverse momentum to
the final partons and the time-like branchings produce
a lot of partons with small $p_T$. The final $p_T$
spectrum then looks more or less like an exponential one,
even though secondary scatterings have not yet
been taken into account.

	In Fig.{}~\ref{fig6}, we plot $dN/dz$ as a function of $z$ at
different time to illustrate how the parton production
evolves in space and time. At $t=-0.7$ fm/c, as the two nuclei
approach toward each other before they actually overlap,
initial state radiations have already begun. These partons
have large rapidities and are Lorentz contracted with an
average spread in $z$,
\begin{equation}
	\Delta z\approx 1/p_0+2R_A\frac{2m_N}{\sqrt{s}}
	\approx 0.25 {\rm fm}, \ \ \sqrt{s}=200 {\rm GeV},
\end{equation}
where $p_0=2$ GeV/c is the $p_T$ cutoff for semihard
scattering, $R_A$ is the nuclear radius of $Au$ and
$m_N$ is the nucleon mass.  After the hard scatterings
and during the interaction time, partons are produced
uniformly in the central rapidity region. Afterwards,
partons follow straight line by free-streaming and are
distributed evenly in $z$ between two receding pancakes
of beam partons (partons from initial state radiations).
One can clearly see that there is approximately boost
invariance in the central region in this free-streaming
picture.

	Another illustrative way to study the evolution
of local parton density is to make a contour plot of parton
density $\rho$ in $z$ and $t$ as shown in Fig.{}~\ref{fig7}.
Here, $\rho$ is defined as
\begin{equation}
	\rho=\frac{1}{\pi R_A^2}\frac{dN}{dz}, \label{eq:rho1}
\end{equation}
and a sharp sphere distribution is assumed for nuclear density.
One can clearly see that partons inside the two approaching nuclei
have a spatial spread of $\Delta z=0.25$ fm in $z$. This spread
continues for partons from the initial
state radiation as they escape from the interaction
region along the beam direction with large rapidities. The
interaction region where semihard scatterings happen lasts
for about 0.5 fm/c, from $t=-0.25$ to 0.25 fm/c.  Because
of the contracted distribution, some of the partons lie
outside the light-cone which is defined with respect to
the overlapping point of the two nuclei.  The definition
of a proper time $\tau$ (Eq.{}~\ref{eq:tau}) for the evolution
of the whole system is therefore only valid within an accuracy
of $\Delta z$.

	If one assumes boost invariance \cite{BJOR},
the parton density can be estimated as
\begin{equation}
	\rho=\frac{1}{\pi R_A^2\tau}\frac{dN}{dy},
\end{equation}
where $\tau$ is the proper time and $\rho$
should be a function of $\tau$ only.  By comparing the
contour of constant density with the hyperbola of
constant $\tau$ (dot-dashed line) in Fig.{}~\ref{fig7},  we see that
this is true only approximately.
We also see that the density, as indicated by the numbers,
decreases like $1/\tau$ due to free-streaming.

\subsection{Local Isotropy in Momentum Space}

	Since there are numerous partons produced within a
rather short time as we have demonstrated, the initial parton
density is very high at $t\ge 0.25$ fm/c, immediately after
the interaction region (see Fig.{}~\ref{fig7}). Within such a dense
system, secondary parton scatterings and production are
inevitable. The equilibration time for the system can be
estimated by solving a set of rate equations as recently has
been done in Ref. \cite{BDMTW}. In this approach, one must
make sure that there is
approximately local isotropy in momentum space. This
could be achieved through secondary parton scatterings
as has been investigated in Refs. \cite{KGBM,SHUR}. One
can also use free-streaming to estimate the upper bound
of the thermalization time $t_{\rm iso}$ \cite{BDMTW,HWA},
by studying the momentum distribution of partons in a
cell of the size of the mean-free-path $\lambda_f$. We
take $\lambda_f\sim 1$ fm in our study here.

	Let's concentrate on the central slice at $z=0$ with
$|z|<0.5$ fm. At the very early stage during the interaction
region (for example at $t=0$ in Fig.{}~\ref{fig7} ), produced
partons with different rapidities are confined
to a highly compressed slab with $\Delta z \approx 0.5$ fm.
As the system expands, partons with large rapidities will escape
from the central slice while partons with small rapidities
remain. As shown in Fig.{}~\ref{fig8}, the rapidity
distribution of the  produced
partons in this central slice evolves from a plateau-like
distribution with width $\Delta y/2\approx 2.5$ at early times
to a Gaussian shape at $t=0.7$ fm/c. If the free-streaming continues,
the rapidity distribution in this central slice will eventually
become a $\delta$-function and only partons with zero rapidity
remain. The evolution in other cells is similar in their local
frames which have spatial rapidity
\begin{equation}
	\eta =\frac{1}{2}\ln\frac{t+z}{t-z}
\end{equation}

	In Fig.{}~\ref{fig9}, we show the evolution of momentum
distributions  in $p_x$ (solid lines) and $p_z$ (dashed lines)
at different times (indicated by the number on each line).
The $p_x$ distribution should evolve like $p_T$ distribution
as is shown in Fig.{}~\ref{fig5}. The slope of
$p_z$ distribution, however, decreases because partons with large
longitudinal momenta gradually escape from the central region.
At $t=0.7$ fm/c, the slopes of $p_x$ and $p_z$ distributions become
the same. If one wants to use the parton production we have calculated
so far as an initial condition to study thermal and chemical
equilibration, this should be the starting point where local
isotropy is approximately achieved. Therefore, the time
for achieving isotropy in momentum space via free-streaming is
\begin{equation}
	t_{\rm iso}\approx 0.7 \ \ {\rm fm/c}.
\end{equation}
This is the same as estimated in Ref. \cite{BDMTW}. We emphasize
that this is only an estimate of the upper bound. However, without
secondary scatterings, thermalization can never be achieved and
maintained.

\subsection{Chemical Composition}

	Unlike an ideal gas of quarks and gluons in which chemical
equilibrium is maintained, the initial production of quarks and
gluons is determined by the parton structure functions, the hard
scattering cross sections and the radiation processes in pQCD.
Due to the difference in the numbers of degrees of freedom in
color space, cross sections involving gluons are always larger
than those of quarks. For small angle scatterings, one can
verify that,
\begin{equation}
	\frac{d\sigma_i}{dt}\cong C_i \frac{2\pi\alpha_s^2}{t^2},
\end{equation}
where $t$ is one of the Mandelstam variables and,
\begin{equation}
	C_i=\frac{4}{9}, 1, \frac{9}{4},
\end{equation}
for $i=qq$, $gq$ and $gg$ scatterings.
Similarly, both initial and final state radiations produce
more gluons than quarks and anti-quarks. Therefore in pQCD,
the ratio between produced quarks and gluons is much smaller
than the ratio of an ideal gas, which is $9/4$ for three quark
flavors.
	Shown in Fig.{}~\ref{fig10} are the fractions of produced
quarks and anti-quarks as functions of time $t$ in $pp$ and $AA$
collisions at $\sqrt{s}=200$ GeV/n. For $Au+Au$ collisions,
hard scatterings with $p_T>2$ GeV/c produce about 13\% quarks
and anti-quarks. If initial and final state radiations are
not included the ratio jumps to 18\%, because radiations
produce more gluons than quarks and antiquarks. In $pp$ collisions,
about 28\% of the partons produced via hard scatterings without
radiations are quarks and anti-quarks.
	The difference between $Au+Au$ and $pp$ collisions is
due to the different $A$ dependence of the valence and sea quark
production. Since the partons are produced via binary collisions,
the number of produced gluons and sea quarks and antiquarks scales like
$A^{4/3}$. On the other hand, baryon  number conservation requires
valence quark production to scale like $A$. Given the fraction of
total quark $q_{pp}=0.28$ and valence quark $v_{pp}=0.14$
production in $pp$ collisions, one can find for $Au+Au$ collisions,
\begin{equation}
	q_{AA}=\frac{q_{pp}-(1-A^{-1/3})v_{pp}}{1-(1-A^{-1/3})v_{pp}},
\end{equation}
which gives $q_{AA}=0.18$ for $A=197$ as we obtained from
Fig.{}~\ref{fig10}  of the numerical calculation. To demonstrate the
effective valence quark production, we plot in Fig.{}~\ref{fig11} the
rapidity dependence of the fractional quark number.
In $pp$ collisions, the valence quark production peaks at
large rapidities. In nucleus-nucleus collisions, the valence
quark production has a different scaling in $A$ than gluons and
sea quarks. This is why the fraction of quark and antiquark
production is suppressed more at large rapidity than in the
central rapidity region. The relative quark and antiquark
production in the central rapidity region is further
suppressed by final state radiation.

	The small fractional quark production within pQCD
has important consequences for chemical equilibration of
the partonic system. Because of the small initial relative
quark density and small quark production cross section
as compared to gluon production, it takes a very long time,
if ever, for the system to achieve chemical equilibrium.
If this time is longer than the phase transition time,
a fully equilibrated QGP may never be formed \cite{BDMTW}.

\subsection{Consequences on Parton Rescatterings}

	So far we have not considered final state interactions
among the produced partons. Though an exact quantum field
treatment of parton cascading is not possible with present
technology, various semi-classical approximations has been
undertaken \cite{KGBM,RANF}. The most important
and difficult effort in these semiclassical approximations
is to emulate the quantum effects, like interference,
coherence, and especially, nonperturbative phenomena in QCD. Since we
have studied the initial parton production, a discussion
of the consequences on how to treat final state parton
rescatterings seems to be in place now.

	We have explicitly taken into account the interaction
time for the semi-hard scatterings, which is roughly
$t_i\sim 1/p_T$. Inside a highly Lorentz contracted nucleus,
the spatial spread for partons which could participate in a hard
scattering with transverse momentum transfer $p_T$ is also
about $\Delta z\sim 1/p_T$. This would leave the produced
partons no time to have another hard scattering of $p_T$
with the incoming beam partons. For finitely contracted nuclei
at relatively low energies, this
kind of double high $p_T$ scattering is still possible. However,
these kind of higher twist processes should be
suppressed by a factor of $1/p_T^2$. It is also
possible for a parton to go through a hard and a soft
scattering  subsequently since the soft partons always have
a spatial spread of 1 fm.  This kind of hard-soft multiple
interactions constitute the leading contribution to higher
twist corrections to hard processes in nuclear
collisions \cite{QIU}. However, double semihard
scatterings at high energy  with $p_T\sim p_0$ will be
suppressed due to finite interaction time.

	The large $p_T$ enhancement of both Drell-Yan dilepton
production and single hadrons in $pA$ collisions at $\sqrt{s}\leq 50$
GeV is considered as a result of multiple parton interactions
\cite{BROD}. At these energies, the contracted length of a heavy
nucleus is still relatively larger than the interaction and formation
time. The factorized form of parton model is then modified
due to the finite beam energy. However, a collection of
experimental data \cite{CRON1,CRON2} even in this intermediate
energy range has already shown the effects of the finite
interaction time. As the energy increases, the interaction
time becomes more important as compared to the size of an
increasingly contracted nucleus. The partons then have less
time for secondary scatterings. This then leads to the
observed decrease of the large $p_T$ enhacement in $pA$
interactions as the beam energy increases. At ultrarelativistic
energies, one should therefore expect the suppression of double
semihard scatterings.

	In this study, we used a QCD-based probabilistic model
to calculate the number of multiple independent semi-hard parton
scatterings. When calculating the inclusive jet cross section in
Eq.{}~\ref{eq:sjet}, we used the QCD-evolved parton structure
function $f_{a/A}(x,Q^2)$ at the scale of the hard scattering $Q^2$.
Therefore by sampling the partons from a joint distribution
function of two colliding nuclei at $Q^2$,
\begin{equation}
		f_{a/A}(x_1,Q^2)f_{b/B}(x_2,Q^2),
\end{equation}
we have already included the probability for partons from
initial state radiation to participate in the semihard
scatterings. As illustrated in Fig.{}~\ref{fig12}, the two
partons which are chosen to have hard scatterings may originate from
the same initial branching tree. Since not every parton
has to have hard scattering in our probabilistic model,
other partons produced from the initial state radiation
tree have been chosen not to suffer hard scatterings.
These partons should not be allowed to have further
scatterings with the beam partons, even though they
are produced long before the hard scattering and
have to ``go through'' the interaction region during the
space-time evolution (see Fig.{}~\ref{fig7}). Another way to
understand this is to consider all partons in the
initial state radiation tree as fluctuations. During
the hard scatterings, the beam partons act as probes
which can resolve the fluctuations up to scale $Q^2$.
These partons on the branching tree leading to the
hard scatterings will become incoherent, while
those not leading to any hard scattering will
eventually reassemble back to the initiating
partons. In our probabilistic model, the beam
partons will be chosen to interact with only some of
these resolved partons, while others will simply
move along the beam direction even though they become
incoherent to the nucleons.

	In our simulation, partons in different hard
scatterings have their own {\it independent} initial state radiation
trees, decreasing the virtuality from $Q$ backward to
an initial scale $Q_0$ below which everything is considered
nonperturbative. This corresponds to the situation of
$Q_2=Q_0$ in Fig.{}~\ref{fig12}, where both  parton fission and
fusion are included in the nonperturbative regime. However,
the two branching trees could also be correlated as in
Fig.{}~\ref{fig12}
if $Q_1=Q_0$.  In this case, parton fusion and fission
take place in the perturbative regime and the parton structure
functions will be modified \cite{KEXW}. One, therefore, should use
multiple parton correlation function $f(x_1,\ldots,x_i,Q^2)$.
In this paper, we simply used a parametrization \cite{WANG}
to take into account the correlation effect. If this
is not enough, we might have overestimated the parton production
from the initial state radiation.

\section{CONCLUSIONS}

	We have studied the space-time evolution of initial
parton production in ultrarelativistic heavy ion collisions.
At RHIC energy, we found that the production time is about
$0.5$ fm/c after the two colliding nuclei have complete overlap.
At $t=0.7$ fm/c, the produced partons inside a cell $|z|\leq 0.5$ fm
have momentarily achieved local isotropy
in momentum space by free-streaming, consistent with the estimate
in Ref. \cite{BDMTW}. This could provide a starting point if one wants
to investigate the equilibration of the produced partons via
rate equations. The initially produced partons are however far away from
chemical balance due to small quark production cross sections.

	The formation time of the produced partons has serious
consequences on the subsequent secondary scatterings. The number
of double semihard scatterings should be suppressed
if one takes into account the formation time. Due to
factorization, partons from the associated initial state
radiation will also not scatter again with the incoming beam
partons.

	In hadronic collisions, minijets become important only at
above $\sqrt{s}=50$ GeV/c. Soft interactions, however, still
have important contributions to the total cross section
and particle production even at the highest collider energy
presently available. What we have calculated in this paper
is only the parton production through the semihard
processes. We have not included the nonperturbative soft
processes. In HIJING model, the soft interactions have been
modeled as string formation, which carries energy by means
of a color field. Naively, one would think these soft
interactions are still present in the final state cascading.
However, as has been shown in Refs. \cite{BMW,KEMG}, the high partonic
density produced by the semihard scatterings will screen the color
field and make the soft component less important.
What are left over may only be those semihard rescatterings among
the produced partons, with
the infrared cutoff replaced by the screening mass.
Then we would use pQCD to simulate the parton cascading consistently.
We hope the study in this paper will pave our way ultimately
to such a cascading model.

\acknowledgments
Discussions with M. Gyulassy are gratefully acknowledged.
KJE thanks Magnus Ehrnrooth foundation and Suomen Kulttuurirahasto
for partial financial support.
This work was supported by the Director, Office of Energy
Research, Division of Nuclear Physics of the Office of High
Energy and Nuclear Physics of the U.S. Department of Energy
under Contract No. DE-AC03-76SF00098.

\figure{ Illustration of a hard scattering and the associated
branching trees. The values of $x_1$ and $x_2$ are the fractional
energy carried by the in-coming partons. The number for each
intermediate line is the life-time of that virtual partons
$\Delta t=q_0/q^2$ (fm), and the numbers in the parentheses
are the corresponding virtualities $q$ (GeV). \label{fig1}}

\figure{The overlap region in space and time of two incoming
partons each with spatial spread of $\Delta z_1$ and $\Delta z_2$,
respectively. The interaction point is chosen randomly inside
the shaded region. Solid lines show the trajectories of the
parent nucleons. \label{fig2}}

\figure{The total number of produced partons $N$ as a function of
time $t$, with $t=0$ defined as when the two nuclei have complete overlap.
 \label{fig3}}

\figure{The rapidity distribution $dN/dy$ of produced partons at different
times $t$ (as indicated by the number for each line). \label{fig4}}

\figure{$p_T$ distributions of produced partons at different times.
	\label{fig5}}

\figure{Parton distribution along the $z$-axis at different times.
	\label{fig6}}

\figure{Contour plot in $z-t$ plane of the parton density $\rho$ of
	Eq.{}~\ref{eq:rho1}, as indicated by the numbers. The wavy
	structure along the light-cone is only an artifact of the
	plotting program. \label{fig7}}

\figure{Rapidity distributions of partons in the central slice of
$|z|<0.5$ fm at different times. \label{fig8}}

\figure{$p_x$ and $p_z$ distributions at different times for
	partons in the central slice $|z|<0.5$ fm. \label{fig9}}

\figure{The fractional number of produced quarks and anti-quarks
as a function of time for $Au+Au$ (solid)collisions, $Au+Au$
without radiations (dashed), and $p+p$ without
radiation (dot-dashed). \label{fig10}}

\figure{The final fractional number of produced quarks and anti-quarks
as functions of rapidity for $Au+Au$ (solid)collisions, $Au+Au$
without radiations (dashed), and $p+p$ without
radiation (dot-dashed). \label{fig11}}

\figure{An illustration of the possible correlation between two
initial branching trees leading to two hard
scatterings. The two branching trees can result from parton splitting
(fission) and recombination (fusion). The boxes represent hard
processes with momentum scale $Q$.
\label{fig12}}


\begin{references}
\bibitem{KAJA} K.{}~Kajantie, P.{}~V.{}~Landshoff and J.{}~Lindfors,
	Phys. Rev. Lett. {\bf 59}, 2517 (1987); K.{}~J.{}~Eskola,
	K.{}~Kajantie and J.{}~Lindfors, Nucl. Phys. {\bf B323}, 37 (1989).
\bibitem{JBAM}J.P. Blaizot, A.H. Mueller, Nucl. Phys. {\bf B289}, 847 (1987).
\bibitem{WANG} X.-N. Wang and M.{}~Gyulassy, Phys. Rev. D {\bf 44},
	 3501 (1991); Phys. Rev. D {\bf 45}, 844 (1992).
\bibitem{KGBM}K. Geiger and B. M\"{u}ller, Nucl. Phys. {\bf B369}, 600(1992);
	K. Geiger, Phys. Rev. D {\bf 47}, 133 (1993).
\bibitem{RANF}I. Kawrakow, H.-J. M\"{o}hring, and J. Ranft,
	Nucl. Phys. {\bf A544}, 471c (1992).
\bibitem{COLL}J. Collins, D. E. Soper, and G. Sterman, Nucl. Phys.
	{\bf B261}, 104 (1985).
\bibitem{PYTH} T.{}~Sj\"{o}strand and M.{}~van Zijl, Phys. Rev. D {\bf 36},
	2019 (1987); T.{}~Sj\"{o}strand, Comput. Phys. Commun. {\bf 39},
	347 (1986); T.{}~Sj\"{o}strand and M.{}~Bengtsson, {\em ibid.}
	{\bf 43}, 367 (1987).
\bibitem{BDMTW}T. B. Bir\'{o}, E. van Doorn, B. M\"{u}ller, T. H. Thoma,
	and X. N. Wang, Duke University preprint DUKE-TH-93-46,
	Phys. Rev. C in press.
\bibitem{KEMG}K.{}~J.{}~Eskola and M.{}~Gyulassy, Phys. Rev. C {\bf 47},
	2329 (1993).
\bibitem{EHLQ}E. Eichten, I. Hinchliffe, K. Lane, and C. Quigg,
	Rev. Mod. Phys. {\bf 56}, 579 (1984).
\bibitem{ALPA}G. Altarelli and G. Parisi, Nucl. Phys. {\bf B126}, 298 (1977).
\bibitem{CAPE} A.{}~Capella and J.{}~Tran Thanh Van, Z.{}~Phys. C {\bf 23},
        165 (1984).
\bibitem{HEUR} P.{}~l'Heureux, {\em et al.}, Phys. Rev. D {\bf 32},
        1681 (1985).
\bibitem{PANC} G.{}~Pancheri and Y.{}~N.{}~Srivastava, Phys. Lett. B
        {\bf 182}, 199 (1986).
\bibitem{DURA} L.{}~Durand and H.{}~Pi, Phys. Rev. Lett. {\bf 58}, 303 (1987);
        Phys. Rev. D {\bf 38}, 78 (1988).
\bibitem{DIAS} J.{}~Dias de Deus and J.{}~Kwiecinski, Phys. Lett. B
        {\bf 196}, 537 (1987).
\bibitem{GAIS} T.{}~K.{}~Gaisser and F.{}~Halzen, Phys. Rev. Lett. {\bf 54},
        1754 (1985).
\bibitem{WANG91a}X.{}~N.{}~Wang, Phys. Rev. D {\bf 43}, 104 (1991).
\bibitem{WANG92} X.-N. Wang, Phys. Rev. D {\bf 46}, R1900 (1992);
                D {\bf 47}, 2754 (1993).
\bibitem{ESKO}K. J. Eskola, LBL preprint LBL-32339,1992,
	Nucl. Phys. {\bf B} in press.
\bibitem{FIED}See e.g. R. D. Field, Applications of Perturbative
	QCD, {\sl Frontiers in Physics}, Vol. 77 (Addison-Wesley,
	1989).
\bibitem{ODOR}R. Odorico, Nucl. Phys. {\bf B172}, 157 (1980).
\bibitem{WEBB}G. Marchesini and B. R. Webber, Nucl. Phys. {\bf B238},
	1 (1984).
\bibitem{AM88}A. Mueller, Nucl. Phys. {\bf A498}, 41c (1988).
\bibitem{DKOW}D. W. Duke and  J. F. Owens, Phys. Rev. D {\bf 30}, 50 (1984).
\bibitem{BJOR}J. D. Bjorken, Phys. Rev. D {\bf 27}, 140 (1983).
\bibitem{SHUR}E. Shuryak, Phys. Rev. Lett. {\bf 68}, 3270 (1992).
\bibitem{HWA}R. C. Hwa and K. Kajantie, Phys. Rev. Lett. {\bf 56}, 696 (1986).
\bibitem{QIU}J. Qiu and G. Sterman, Nucl. Phys. {\bf B353}, 105 (1991).
\bibitem{BROD}G.{}~T.{}~Bodwin, S.{}~J.{}~Brodsky and G.{}~P.{}~Lepage,
	Phys. Rev. D {\bf 39}, 3287 (1989).
\bibitem{CRON1}P.{}~Bordalo {\em et al.}, Phys. Lett. B {\bf 193}, 373 (1987);
	D.{}~M.{}~Alde {\em et atl.}, Phys. Rev. Lett. {\bf 64}, 2479 (1990).
\bibitem{CRON2}P.{}~B.{}~Straub {\em et al.}, Phys. Rev. Lett. {\bf 68},
	452 (1992).
\bibitem{KEXW}K. J. Eskola, J. Qiu and X.-N. Wang, LBL preprint, LBL-34163.
\bibitem{BMW}T. Bir\'{o}, B. M\"{u}ller, and X.-N. Wang,
	Phys. Lett. {\bf B283}, 171 (1992).

\end{references}
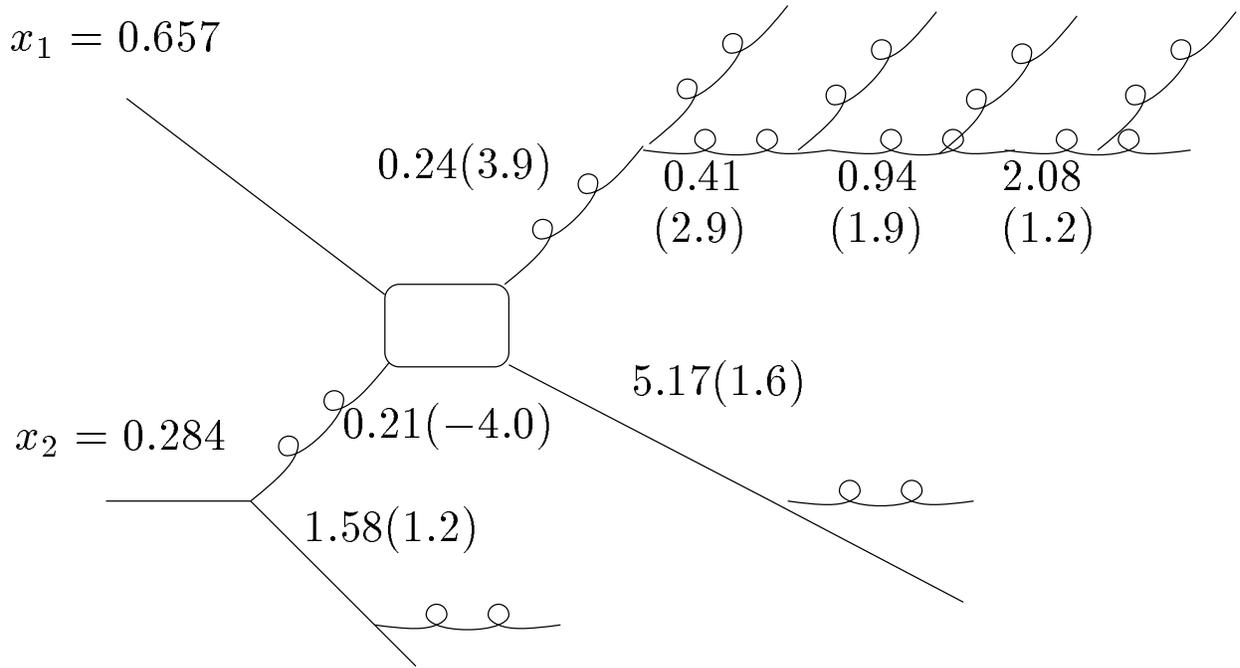
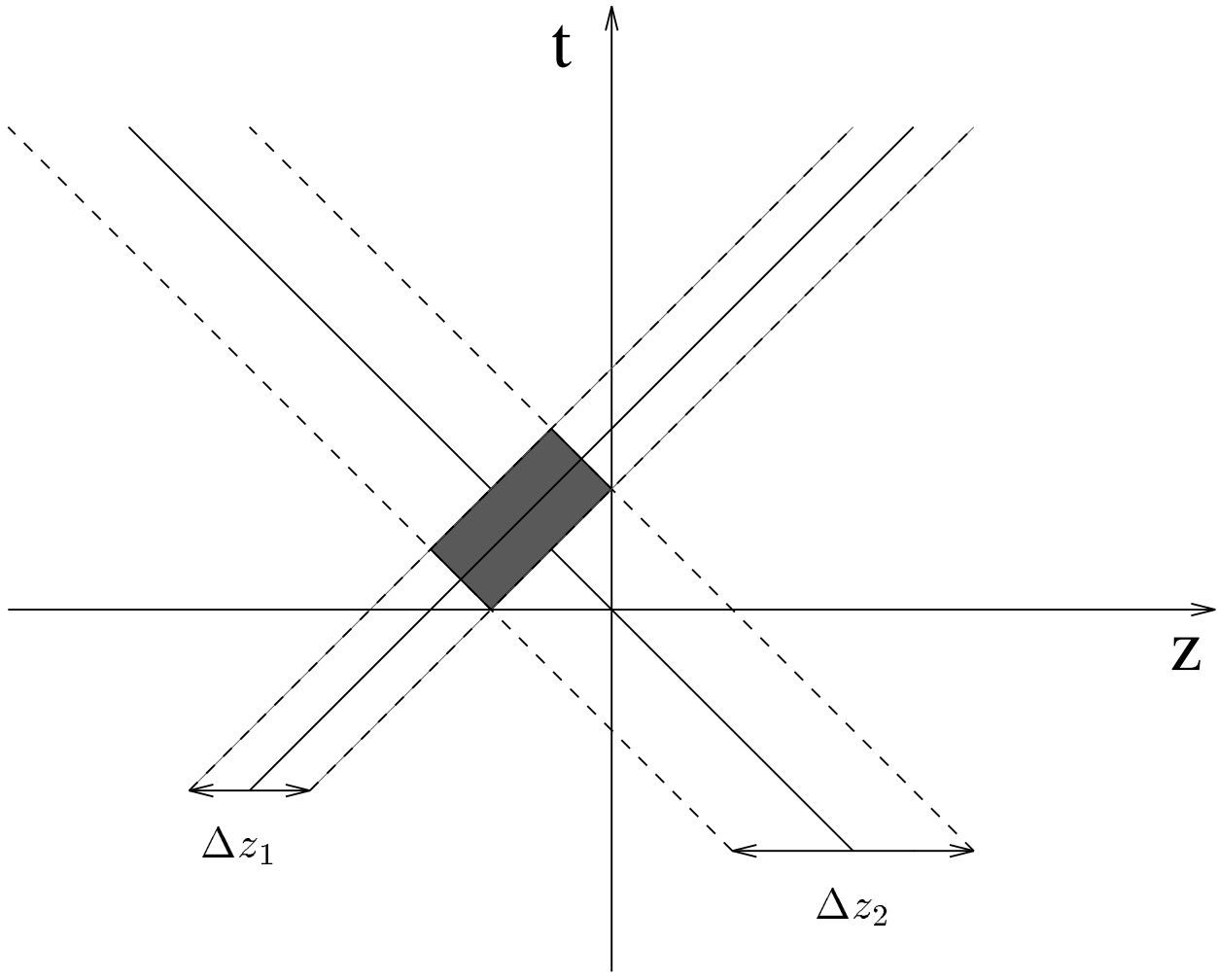
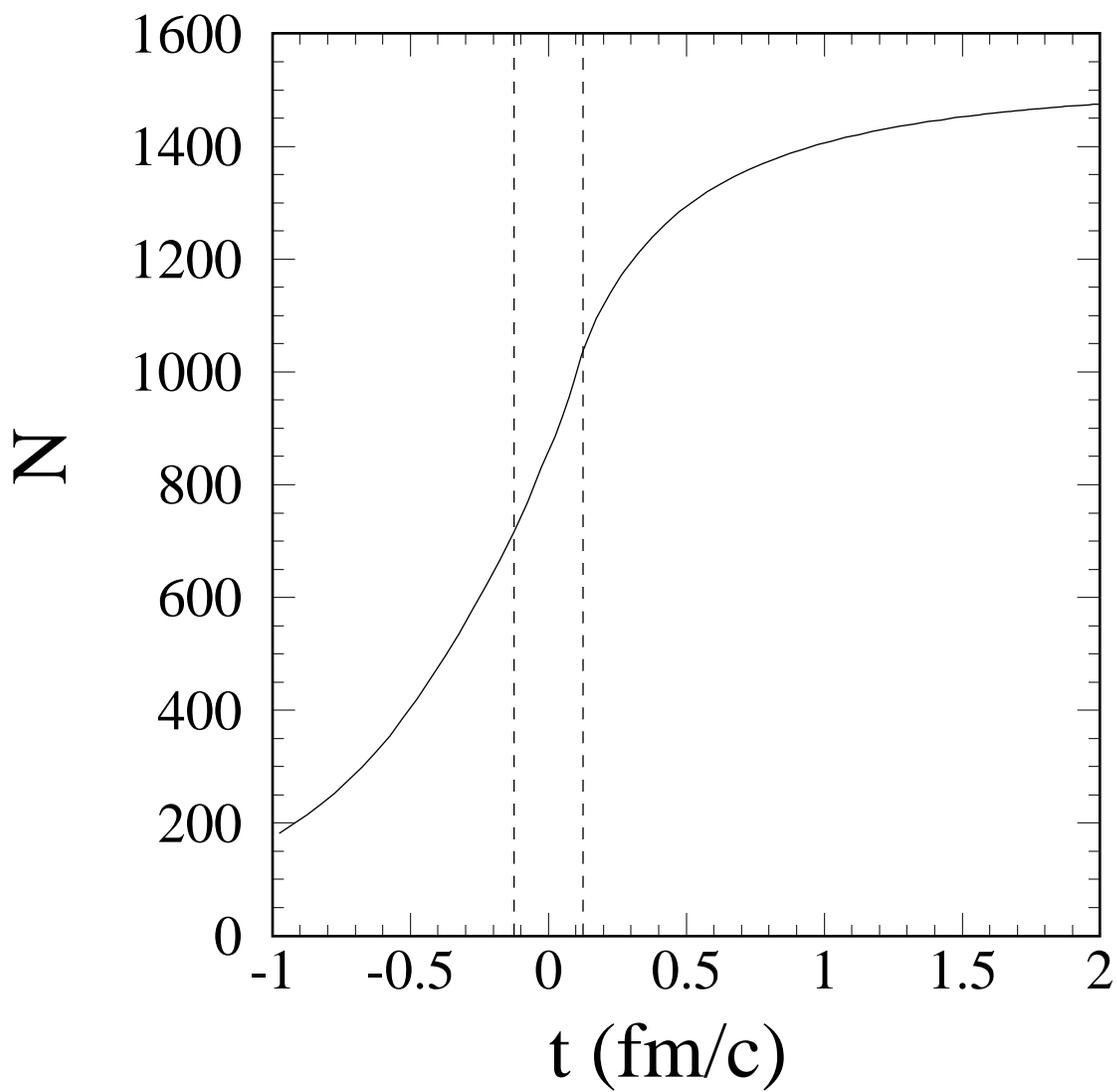
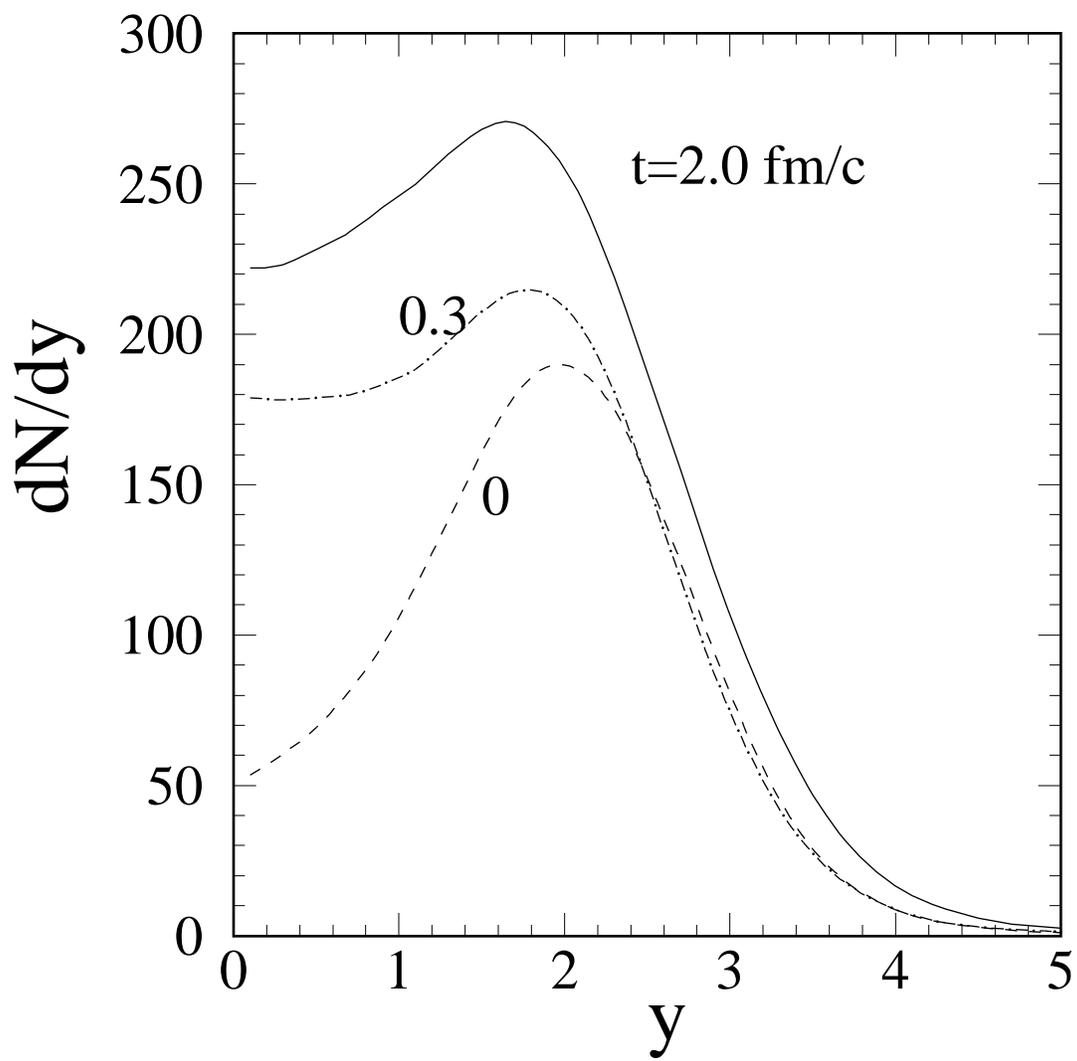
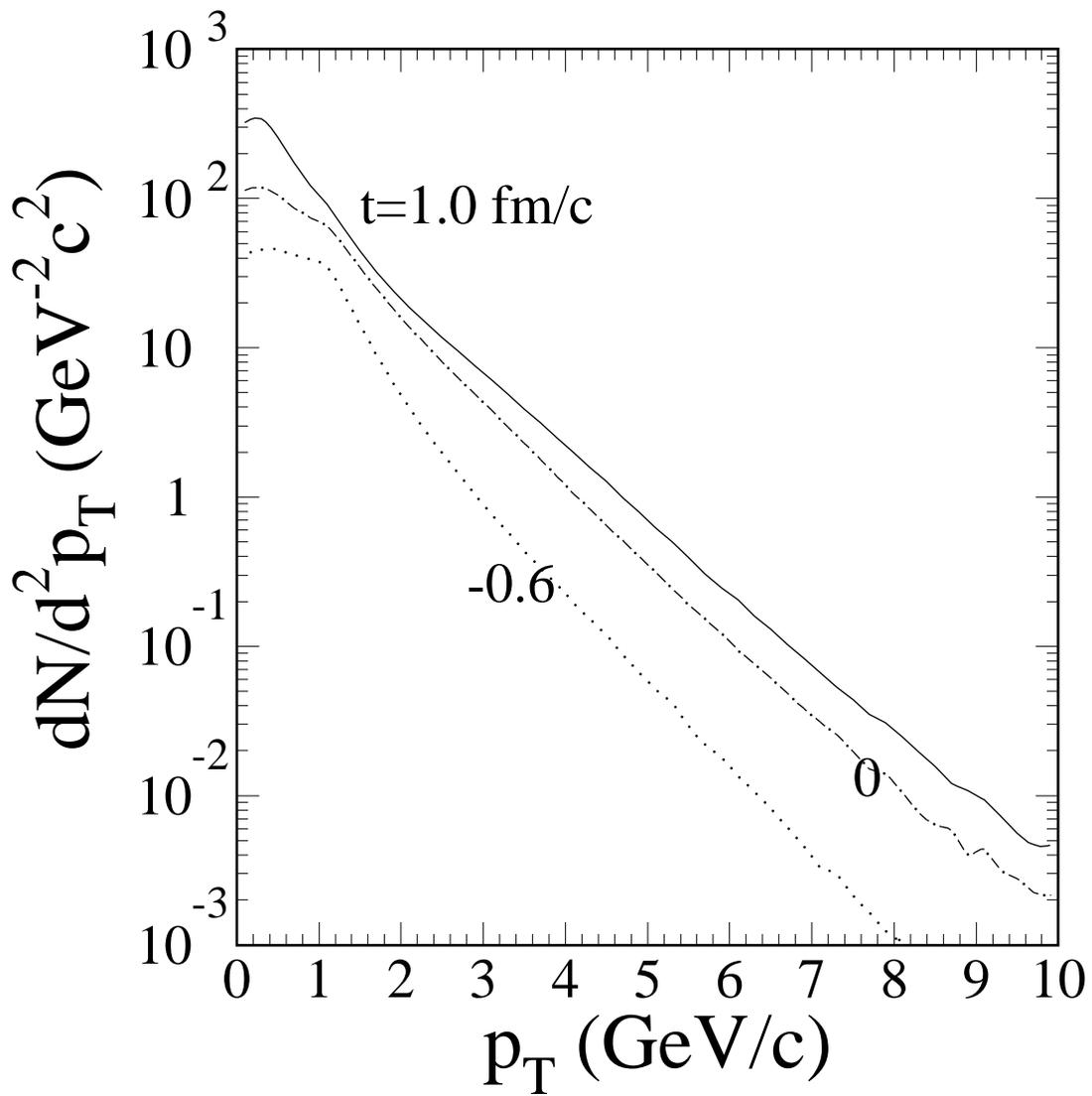
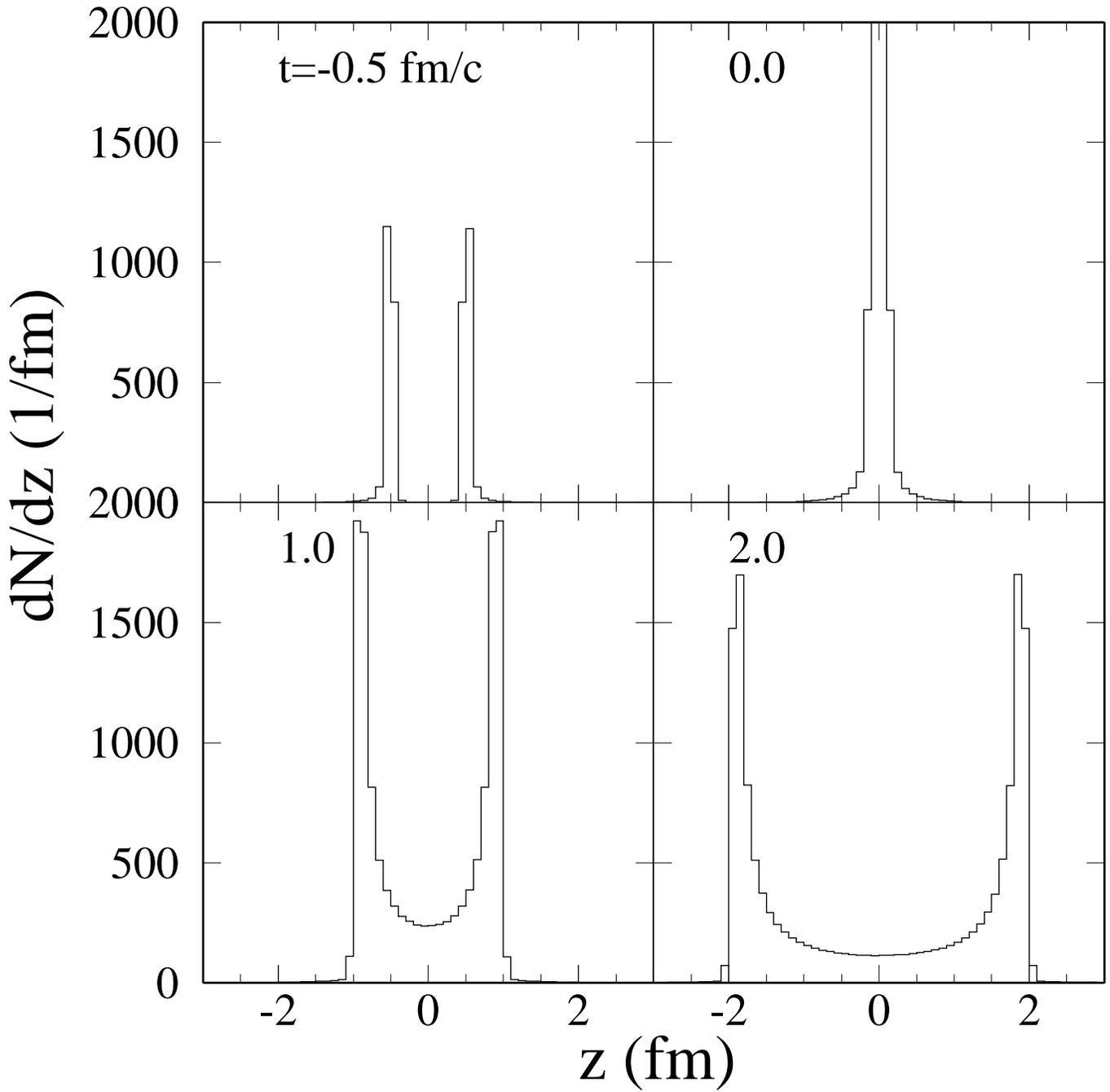
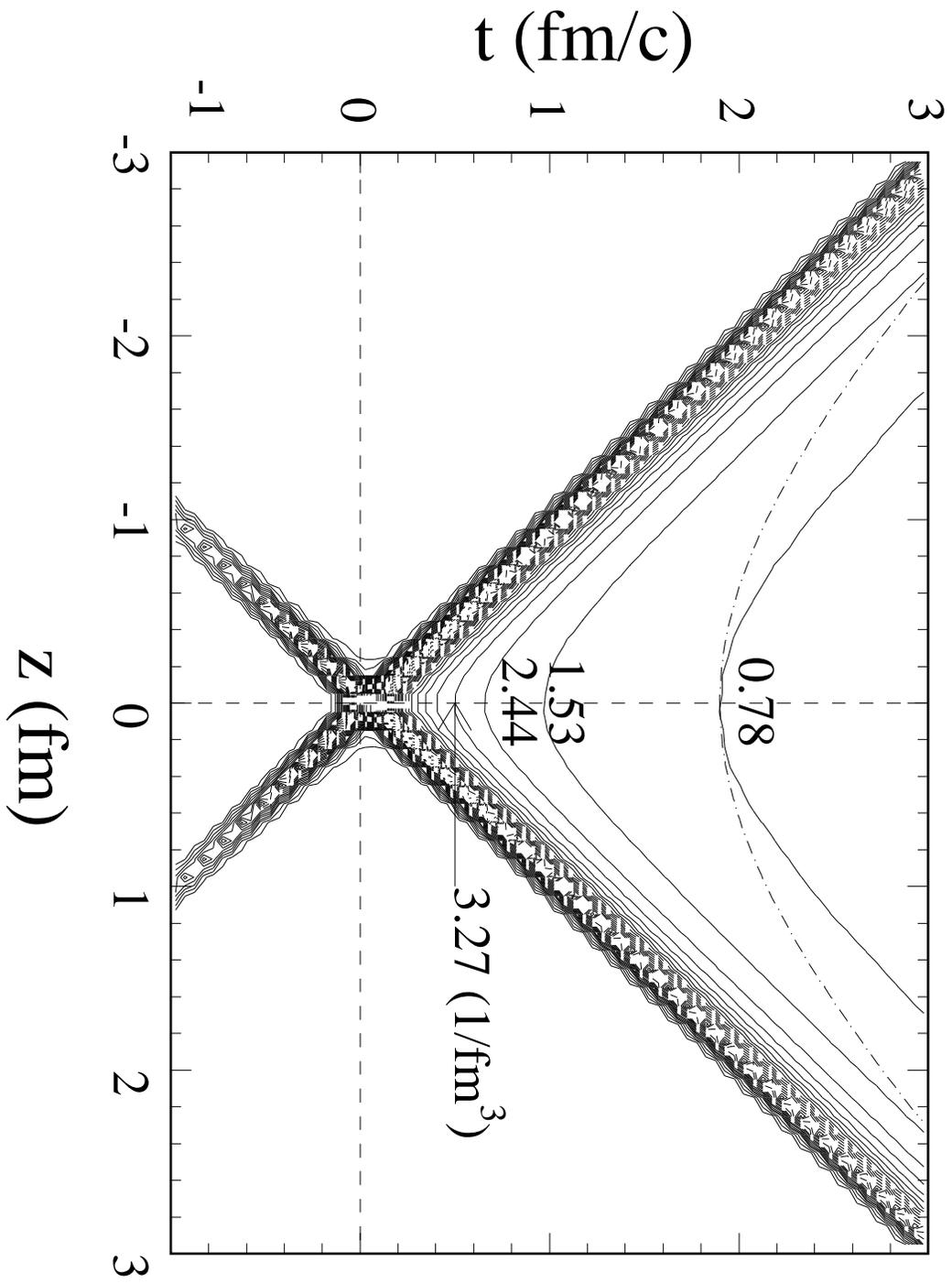
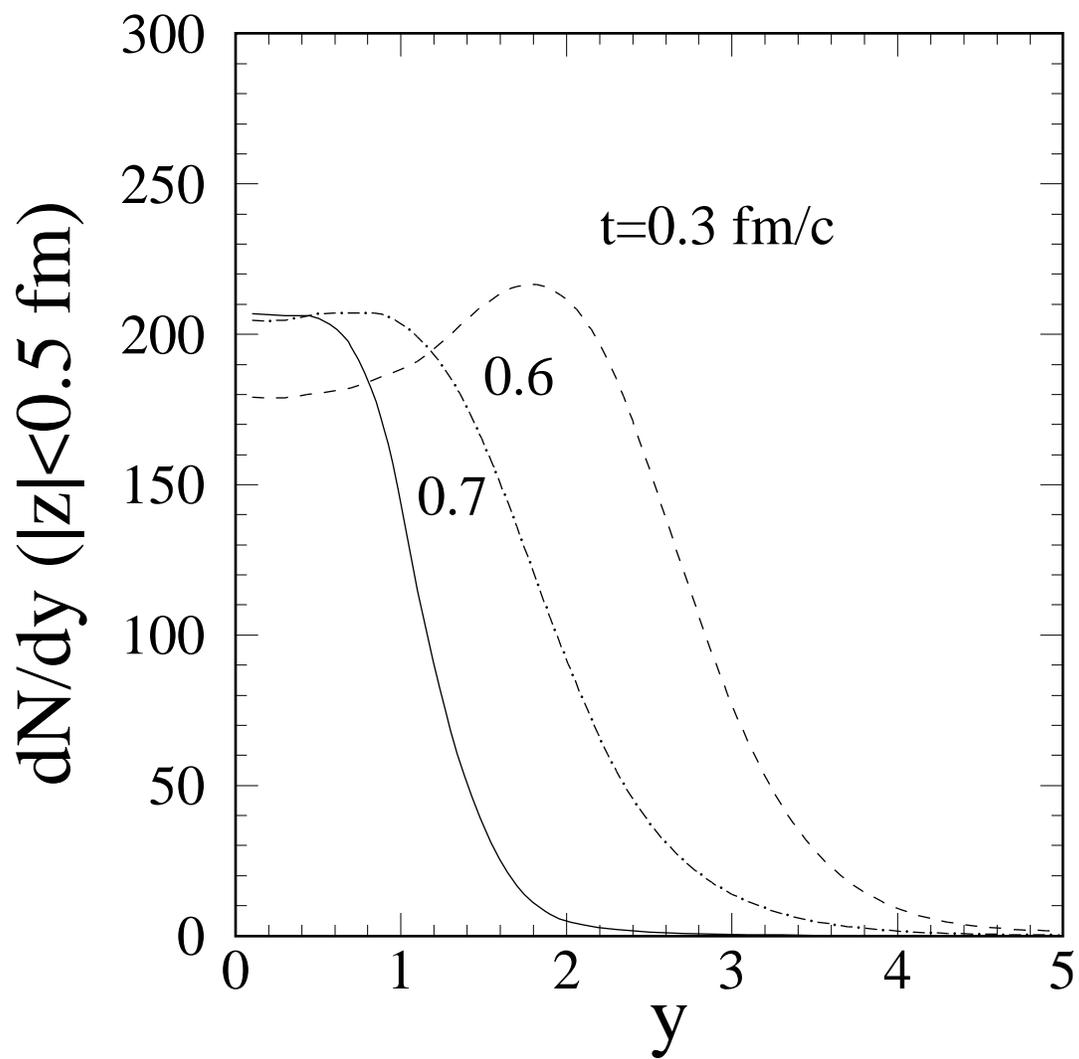
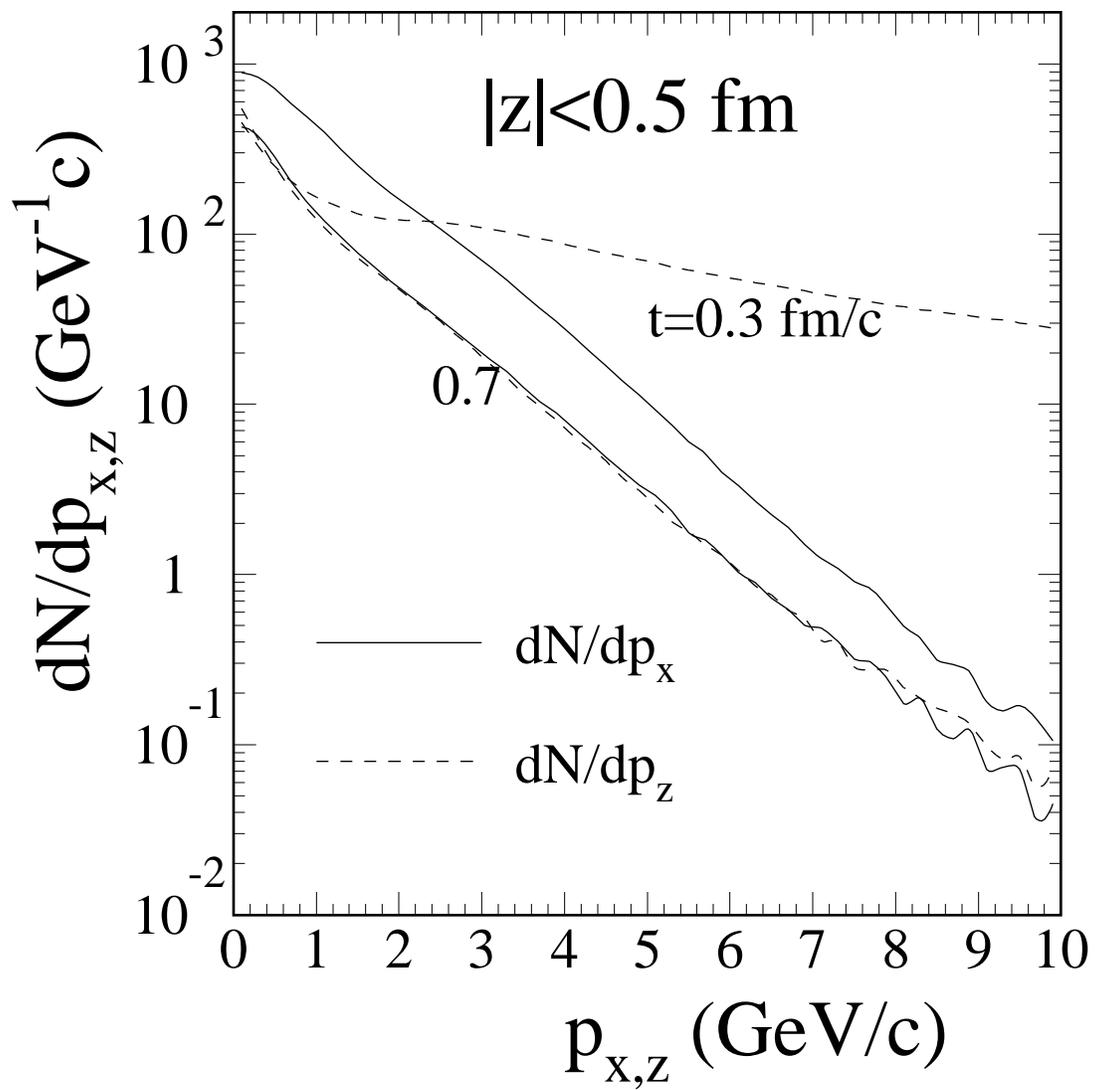
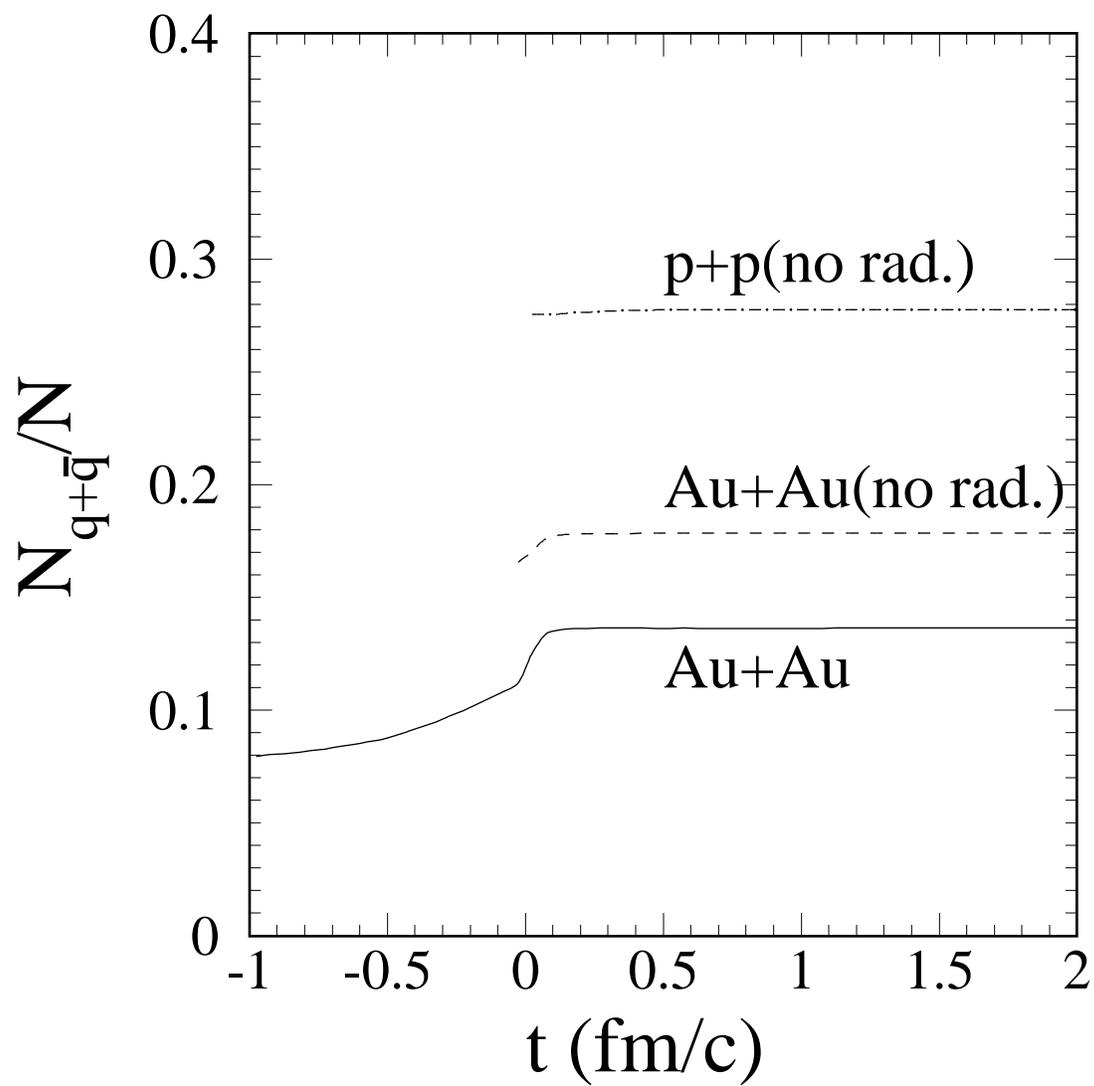
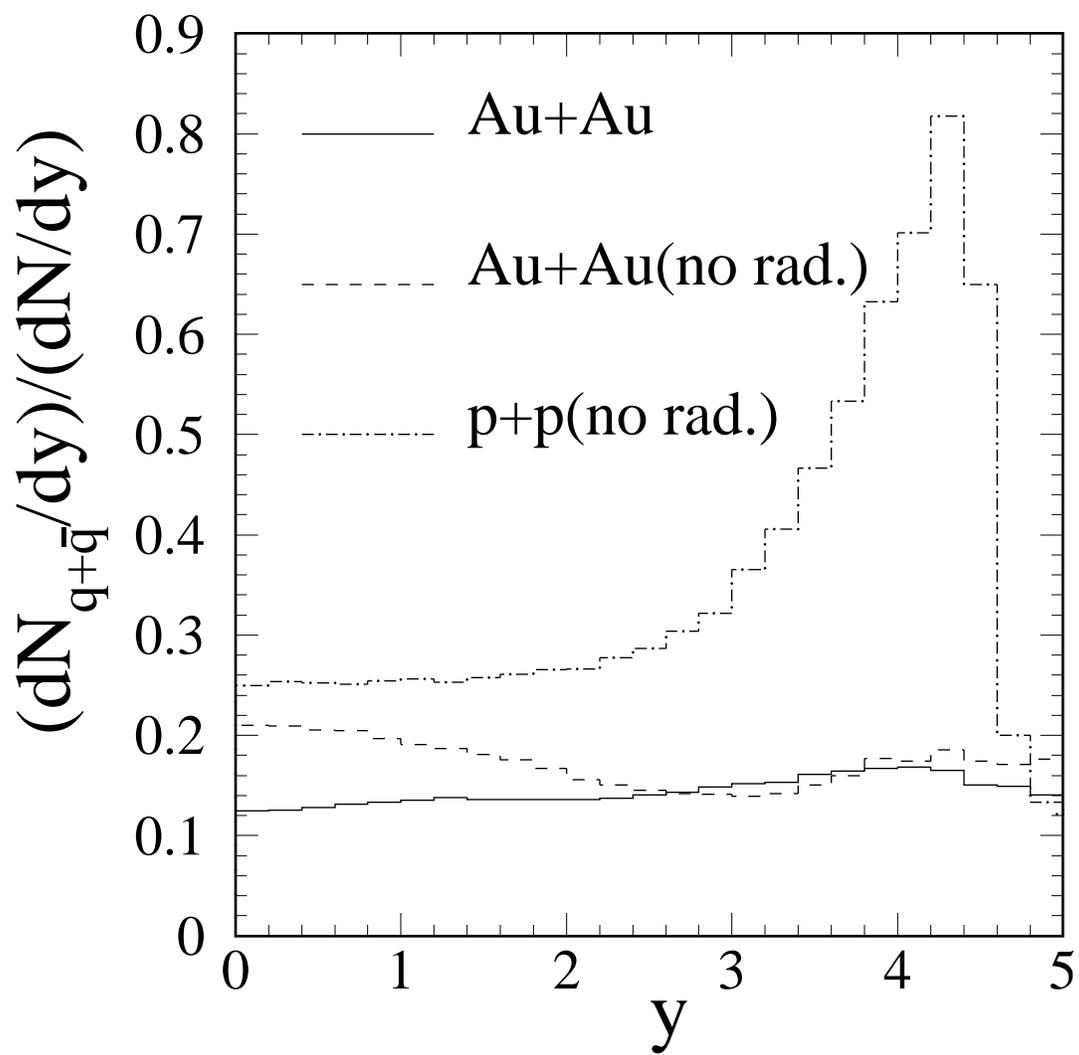
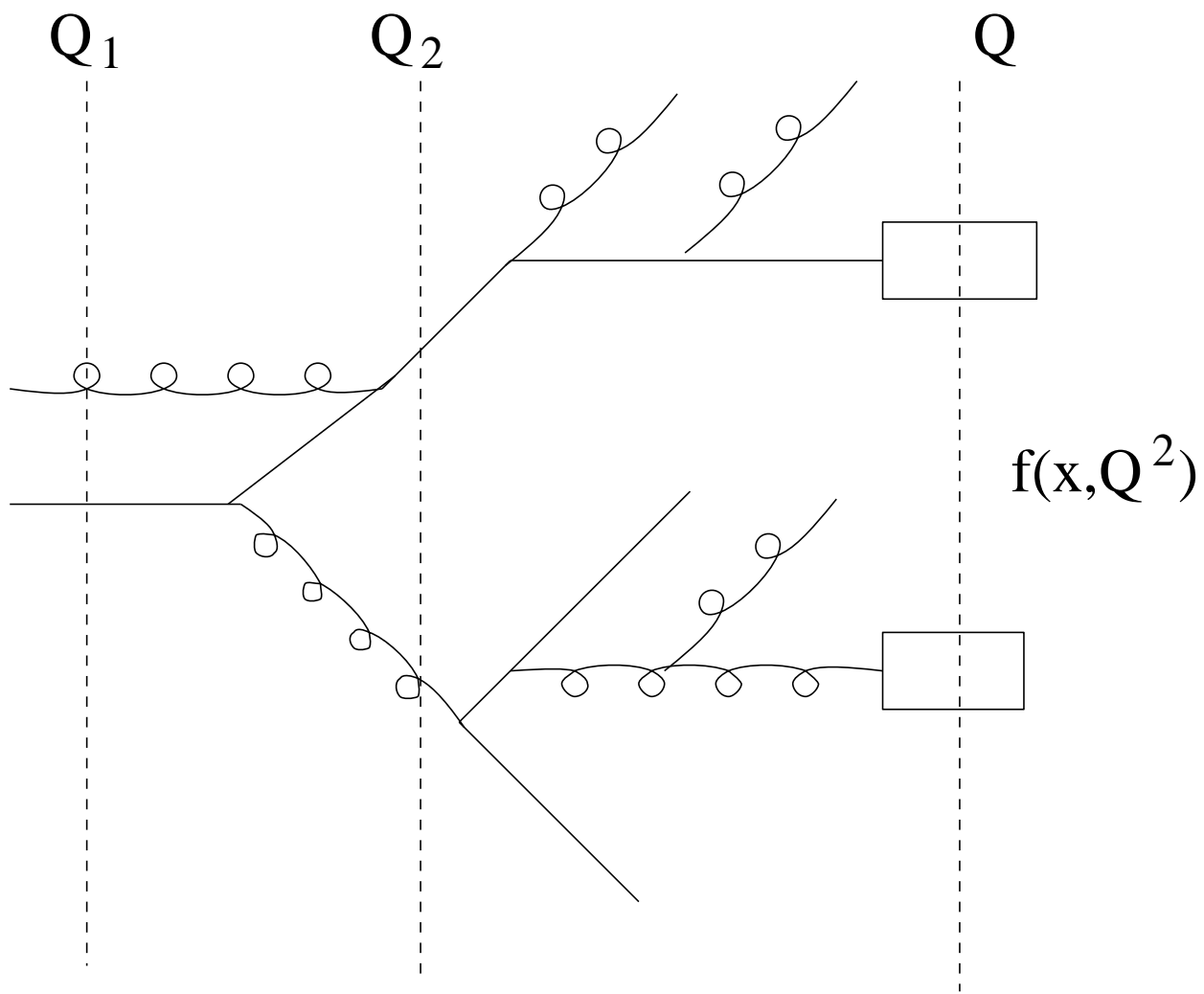
\end{document}